\documentclass[twoside]{procinasan}

\input{procinasan.def}
\usepackage{mathtools}
\usepackage{hyperref}
\usepackage[export]{adjustbox}
\usepackage{subcaption}
\usepackage[section]{placeins}
\usepackage{array}
\newcolumntype{C}[1]{>{\centering\arraybackslash}p{#1}}
\usepackage{booktabs}

\makeatletter

\newcommand{\Rmnum}[1]{\expandafter\@slowromancap\romannumeral #1@}
\makeatother

\begin{document}

\pagebreak

\thispagestyle{titlehead}

\setcounter{footnote}{0}
\setcounter{equation}{0}
\setcounter{section}{0}
\setcounter{figure}{0}
\setcounter{table}{0}

\selectlanguage{english}

\addcontentsline{toc}{subsection}{{\em Zwicky L.N., Molyarova T.S.\/}
Role of CRP-reactions in chemistry of protoplanetary discs}

\addcontentsline{toce}{subsection}{{\em Zwicky L.N., Molyarova T.S.\/}
Role of CRP-reactions in chemistry of protoplanetary discs}

\markboth{Zwicky \& Molyarova 2021}{Role of CRP-reactions in the chemistry of protoplanetary discs}

\titlen{The impact of rates of reactions with cosmic ray induced photons on chemical composition of protoplanetary discs}
{Zwicky L.N.$^{1}$, Molyarova, T.S.$^2$}
{$^1$Ural Federal University, Yekaterinburg, Russia\\
$^2$Institute of Astronomy, Russian Academy of Sciences, Moscow, Russia}

\abstre{Cosmic rays and cosmic ray induced photons are vital components of chemical evolution in areas of interstellar medium that are impenetrable by external ultraviolet radiation. However, rates of reactions with cosmic ray induced photons used in astrochemical models were calculated for molecular clouds and can be different in protoplanetary discs, where dust grows up to larger sizes. Using ANDES astrochemical model, we study how an increase in both upper dust size and rates of reactions with cosmic ray induced photons can influence species abundances in protoplanetary discs. We show that the increase in these reactions' rates has a significant impact on the ice mass fraction in area between 2 and 20 au but has little impact on ionisation degree in disc.}

\selectlanguage{russian}

\addcontentsline{toc}{subsection}{{\em Zwicky L.N., Molyarova T.S.\/} The impact of rates of reactions with cosmic ray induced photons on chemical composition of protoplanetary discs}

\selectlanguage{english}

\addcontentsline{toce}{subsection}{{\em Zwicky L.N., Molyarova T.S.\/} The impact of rates of reactions with cosmic ray induced photons on chemical composition of protoplanetary discs}

\baselineskip 12pt
\section*{1. Introduction}\label{ch:review}

Cosmic rays play an important role in the interstellar chemistry, as they are one of the main sources of ionisation in the interstellar medium and can penetrate even the densest clouds~\cite{2013ChRv..113.9043V}. Cosmic rays not only ionise matter directly, they also produce ultraviolet radiation (at 75 to 300\,nm) that is called cosmic ray induced photons (hereafter referred to as CRPs). They come from de-excitation of vibrational and electronic states of molecular hydrogen, which is, in turn, excited by secondary electrons from ionisation of the medium by cosmic rays or other secondary electrons~\cite{2013ChRv..113.9043V, 1983ApJ...267..603P}. Ultraviolet (UV) radiation itself plays an important role in chemistry because it is involved in the processes of photodissociation, photodesorption and photoionisation of molecules and atoms~\cite{2013ChRv..113.9043V}. In this case, the photodissociation of a molecule can occur either in the gas phase or on the dust surface. In high-density regions, where the absorption $A_V \gtrsim 5$, for a typical interstellar dust particle size distribution, the UV contribution becomes negligible~\cite{2013ChRv..113.9043V}. In these regions, the main component of the UV radiation field become CRPs.

Apart from molecules, CRPs can also be absorbed by dust. Efficiency of the reactions with CRPs (CRP-reactions hereafter) is determined by a fraction of photons that are available for absorption. This is how dust opacity can influence the impact of CRPs on the chemistry. Molecules naturally contribute to the available CRP budget as well but they absorb significantly less radiation than compared to dust~\cite{1989ApJ...347..289G}.

The work by~\cite{1989ApJ...347..289G} was first to evaluate rates of CRP-reactions which was done for the conditions of dense interstellar clouds. These values are used in many of astrochemical models to this day~\cite{2013A&A...550A..36M, 2012ApJS..199...21W}. 
However, there is evidence of dust growth in protoplanetary discs where it can reach sizes orders of magnitude higher than in dense molecular clouds~\cite{2014prpl.conf..339T}. This growth must lead to a change in rates of CRP-reactions (see Section~2). Dust size impacts other aspects of chemistry as well. For example, it affects rates of surface reactions as they depend on cumulative dust surface and in recombination rates with dust as they depend on dust number density. Moreover, dust properties influence the radiation field in the disc. All these effects must be accounted for when investigating the impact of CRP-reaction rates.

The aim of this work is to evaluate the impact of the change in CRP-reaction rates due to dust growth in protoplanetary discs on their chemical composition. We run chemical simulations for different dust sizes and CRP-reaction rates and look at two generalised characteristics of chemical composition: ionisation fraction and ice mass fraction.

\section*{2. CRP-reaction rates}

According to~\cite{1989ApJ...347..289G}, CRP-reaction rate $R_M$ (cm$^{-3}$s$^{-1}$) of a chemical species $M$ with number density $n(M)$ (cm$^{-3}$) is calculated as follows:
\begin{equation}\label{gredelrates}
R_M = \zeta n(M) \int \frac{\sigma_M(\nu)P(\nu)}{\sigma_g(1-\omega) + \sum_M \kappa_M \sigma_M (\nu)} \, d\nu\, ,
\end{equation}
where $\zeta$ (s$^{-1}$) is cosmic ray ionisation rate, $\sigma_M(\nu)$ (cm$^{-2}$) is photoreaction cross-section, $P(\nu)d\nu$ is probability to emit a CRP in frequency range $d\nu$, $\sigma_g$ (cm$^{-2}$) is dust cross-section, $\omega$ is dust albedo, $\kappa_M = n(M)/n(H)$ is the abundance of the species M..

To simplify Eq.~\ref{gredelrates} we make the following assumptions. First, we neglect the frequency dependence of $\sigma_g$ and adopt an average value. According to Fig. 26 of~\cite{2017A&A...602A.105H}, this value varies only within a factor of 3 for small dust grains. Second, we neglect the contribution of molecules to the absorption ($\sum_M \kappa_M \sigma_M (\nu)$ term in Eq.~\ref{gredelrates}). This assumption is valid for all molecules except CO. However, CO contribution is significant only in very narrow frequency bands, while we consider the frequency-averaged effect. Similar assumptions were adopted by~\cite{1989ApJ...347..289G}.

Under these assumptions, Eq.~\ref{gredelrates} leads to $R_M \propto 1/\sigma_g$. The cross-section of a single dust grain can be estimated as a simple geometric cross-section $\sigma_{\rm mon} = \pi a^2$ as $\lambda \ll a$, where $a$ is the radius of the dust grain~\cite{2010A&A...522A..42S}. Assuming that dust size distribution follows the classic power law $dN\propto a^{-3.5}da$~\cite{1977ApJ...217..425M} between minimum and maximum dust grain sizes $a_{\rm min}$ and $a_{\rm max}$, we get $\sigma_g \propto 1/\langle a \rangle$ where $\langle a \rangle = \sqrt{a_{\rm max} a_{\rm min}}$ is mean dust size. Combining all estimates and assumptions, we see that the CRP-reaction rate is directly proportional to the mean dust size:
\begin{equation}\label{rmprop}
R_M \propto \langle a \rangle.
\end{equation}

\section*{3. Considered models}

We use a two-dimensional quasi-stationary model of a protoplanetary disc ANDES \cite{2013ApJ...766....8A} from~\cite{2017ApJ...849..130M}. In this work, we add a limiting factor on surface reaction rates. The rates are reduced in a way that qualitatively accounts for reactions happening only on the surface and not in the bulk of the ice mantle. 

To investigate the impact of CRP-reaction rates, we vary two parameters: maximum dust size $a_{\rm max}$ and CRP-reaction rate multiplier (a factor by which the reaction rate constant $\alpha$ is multiplied in the chemical network). Values of the parameters in the considered models are presented in Table~\ref{Tab:models}.  

\begin{table}[h!]
\begin{center}
\caption[]{Model set}
\label{Tab:models}

 \begin{tabular}{llll}
  \toprule
Model & Maximum & Average & CRP-reaction\\
 & dust size, $\mu$m & dust size, $\mu$m & rate multiplier\\
  \midrule
Model \Rmnum{1} & $0.25$ & $0.035$ & 1\\
Model \Rmnum{2} & $25$ & $0.35$ & 1\\
Model \Rmnum{3} & $25$ & $0.35$ & 10\\
   \bottomrule
\end{tabular}
\end{center}
\end{table}

Dust properties in model \Rmnum{1} correspond to the classic MRN model~\cite{1977ApJ...217..425M}. In model \Rmnum{2} we increase only the dust size, which is useful to separate the effects unrelated to CRP-reaction rate increase. In model \Rmnum{3} we also adjust CRP-reaction rates according to Eq.~\ref{rmprop}. We choose the 10 times increase of the mean dust size. Increasing the dust size further can violate the condition $\sigma_g \gg \kappa_M \sigma_M$ and the adopted assumptions can become unreasonable. Moreover, the chosen maximum size is consistent with a typical size of collisionally growing dust grains. An order of magnitude larger grains should stop growing due to the bouncing barrier~\cite{2014prpl.conf..339T}.

\section*{4. Results}
To analyse the changes in the chemical composition, we consider two generalised characteristics of the medium: ionisation fraction and ice mass fraction.
\subsection*{4.1. Ionisation fraction}
We calculate ionisation fraction as the ratio of a cumulative number density of all charged particles to the gas number density. Gas-phase chemistry is mainly driven by charged particles (ions and electrons)~\cite{2005pcim.book.....T}, so its activity is highly sensitive to ionisation fraction.

\begin{figure}[t]
\subfloat{
\includegraphics[width=0.5\textwidth]{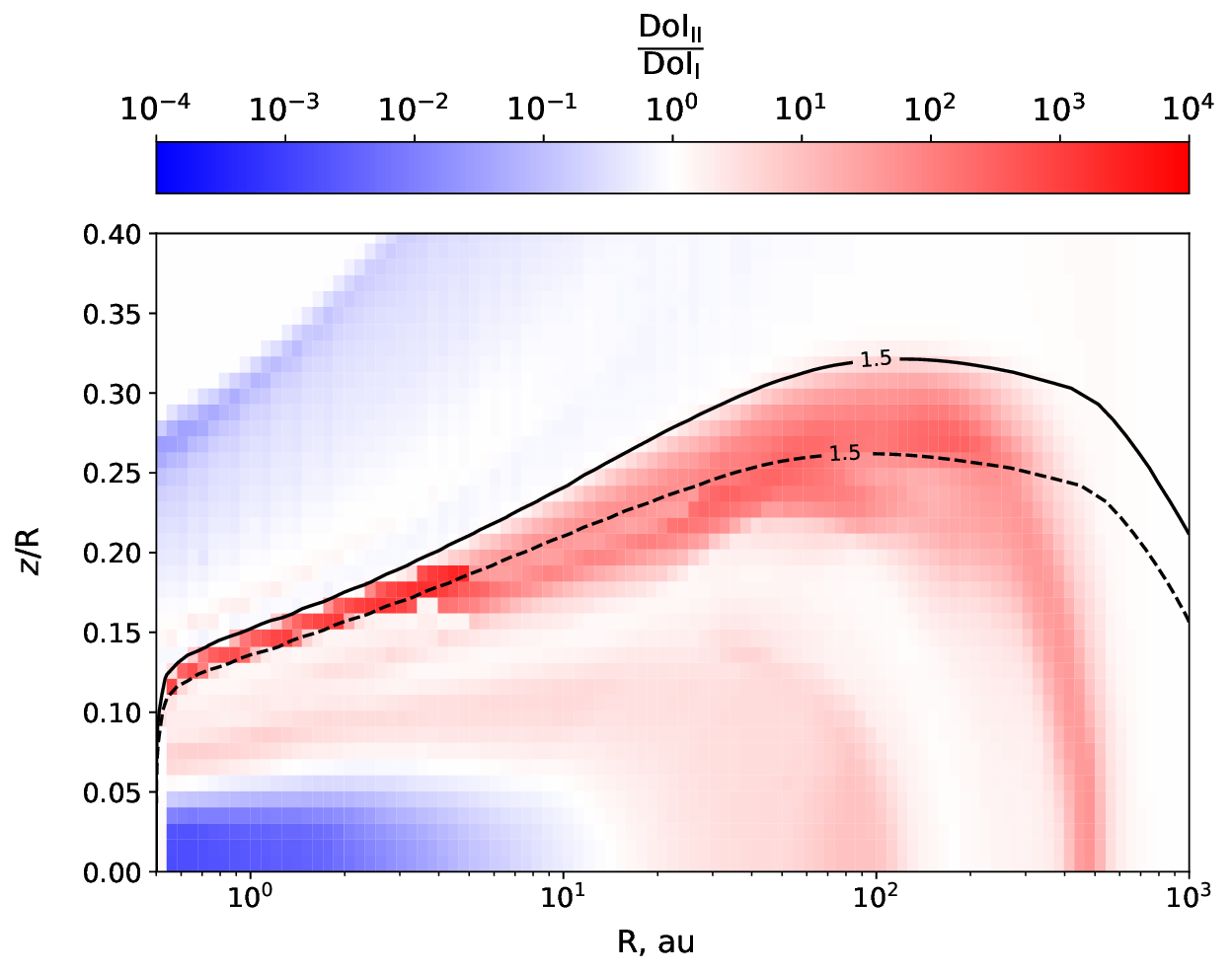}
}
\subfloat{
\includegraphics[width=0.5\textwidth]{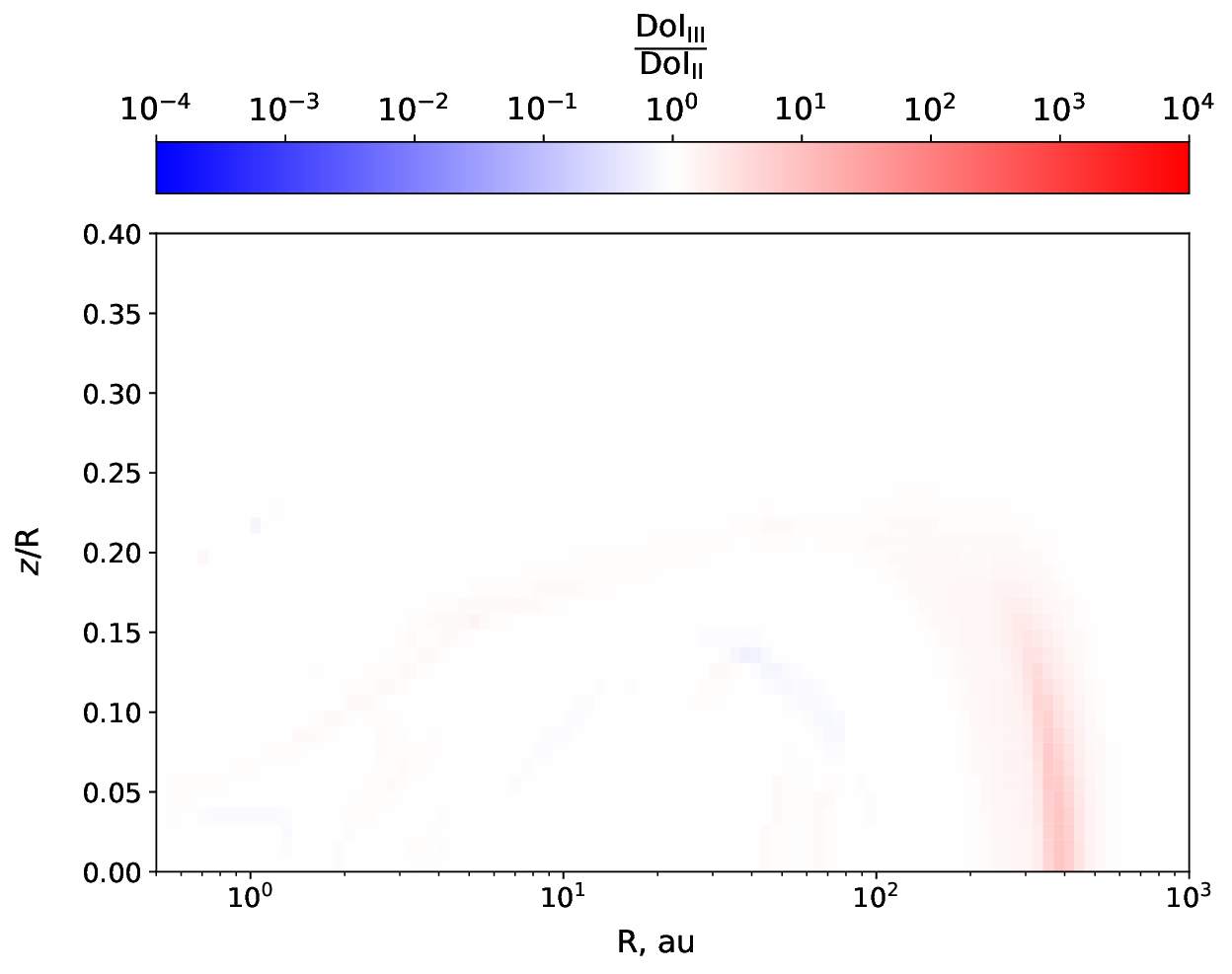}
}
\caption{Distributions of the ratios between different models at the 1\,Myr time moment. Left panel is the ratio between ionisation fraction in the model \Rmnum{2} to model \Rmnum{1}. Right panel is the same but between model \Rmnum{3} and \Rmnum{2}. Black lines in the left panel are $A_V=1.5$ isolines. The solid isoline is from model \Rmnum{1}, dashed is from model \Rmnum{2}.}
\label{fig:ion_yield}
\end{figure}

Fig.~\ref{fig:ion_yield} shows the spatial distribution of the ratios of ionisation fractions in different models: the left panel presents the ratio between models \Rmnum{2} and \Rmnum{1}, the right panel presents the ratio between models \Rmnum{3} and \Rmnum{2}. Here $R$ (au) is distance from the star in the midplane and $z$ (au) is the height above the midplane. Thus the left panel illustrates the effect of the dust size increase and the right panel illustrates CRP-reaction rates increase. In the left panel we can see that a change in dust size leads to a significant change in ionisation fraction: it is increased in regions from 0.5\,au to 10\,au between $z/R = 0.05$ and $0.20$ and from 10 to 500\,au at $z/R < 0.30$. At $R < 10$\,au in the layer between $z/R < 0.05$ and $z/R > 0.20$ ionisation fraction decreases. The former is caused by a decrease in disc opacity due to a lower dust cross-section, while the latter originates from a drop in rates of ionisation of neutral dust by electrons, which is also caused by a lower dust cross-section.

The right panel of Fig.~\ref{fig:ion_yield} shows the increase in CRP-reaction rates does not significantly affect the ionisation fraction. It is expected in the area with $A_{\rm V}<1$ dominated by interstellar UV. Only in a narrow region around $\approx 400$\,au the ionisation fraction is higher because of C$^+$, which is partially ionised by CRPs.

\subsection*{4.2. Ice mass fraction}\label{sect:ices}
Ices are all species that reside on the grain surface. Even though they are the same atoms and molecules as in gas, they are treated as separate species by the model. In our models, all ices are considered volatile, which means they can be both in gas and ice phase. Volatiles play an important role in planet formation as they define the atmospheric composition of forming planets. They also increase mass and size of the grain they reside on which helps it to evolve into a planetesimal and, consequently, a planet~\cite{2014prpl.conf..363P}. Ice species will be further on denoted with the letter ``g'', e.g. gH$_2$O is water ice.

\begin{figure}[h!]
\centering
\includegraphics[width=0.7\textwidth]{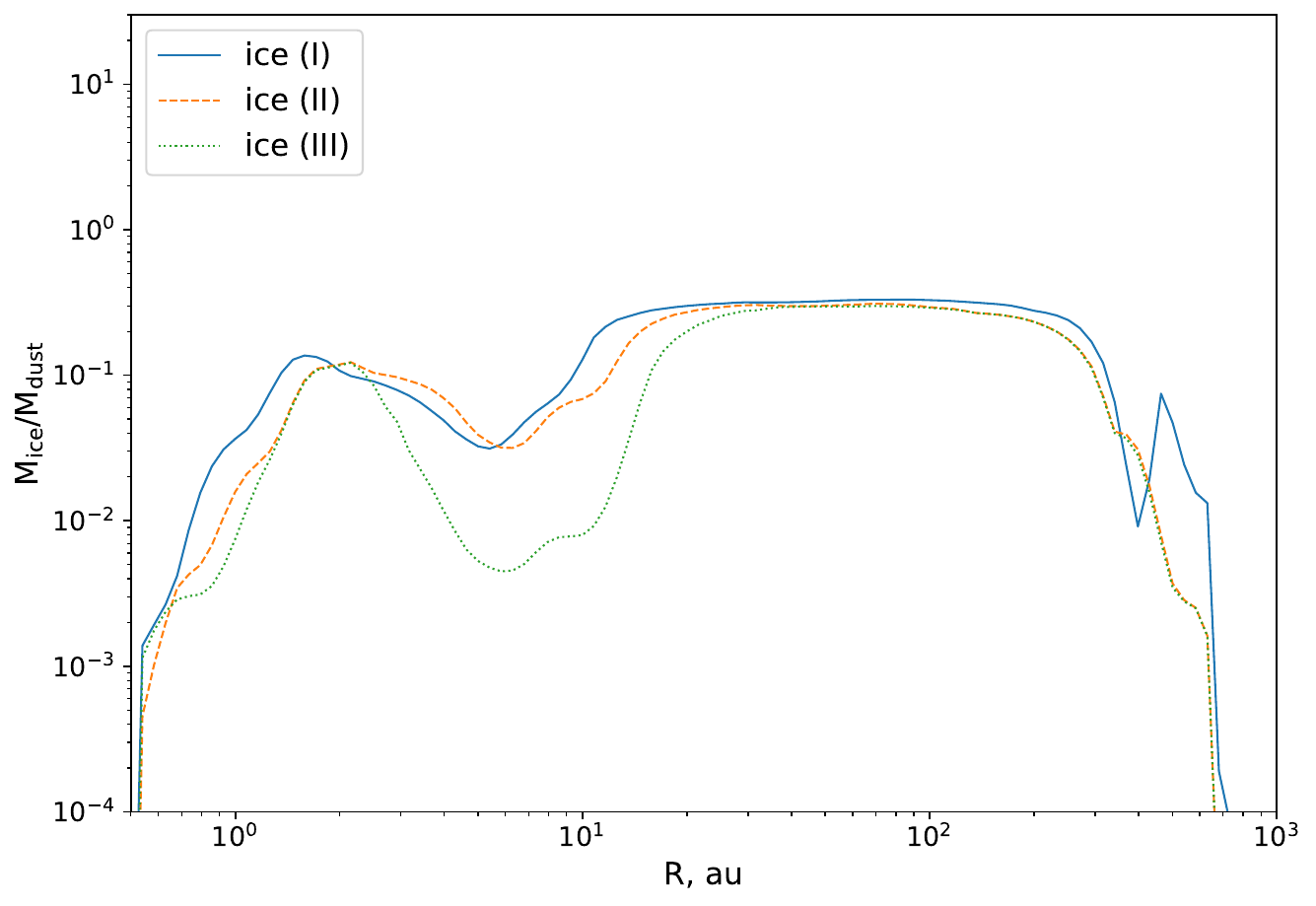}
\caption{Radial distribution of vertically integrated ice mass fraction in different models at the 1\,Myr time moment. Solid line is model \Rmnum{1}, dashed is model \Rmnum{2}, dotted is model \Rmnum{3}.}
\label{fig:icemass1}
\end{figure}

Ice mass fraction is the ratio of cumulative mass of all ice species to the dust mass (M$_{\rm ice}$/M$_{\rm dust}$). Fig~\ref{fig:icemass1} shows radial distribution of ice mass fraction in different models. Dust growth, as can be seen there, leads to a shift of the distribution away from the star and a disappearance of the peak at the disc outer rim. It can be explained by a change in adsorption rate that is proportional to dust cross-section $\sigma_g$, while desoption rate is independent of it. Consequently, their ratio, which defines snowline positions, depends on the dust size. As dust cross-section decreases from model \Rmnum{1} to model \Rmnum{2}, snowlines shift to colder regions. The peak at $\approx 500$\,au in model \Rmnum{1} disappears in model \Rmnum{2} due to an increase in photodestruction rates as disc opacity falls with dust cross-section.

\begin{figure}[h!]
\centering
\subfloat{
\includegraphics[width=0.5\textwidth]{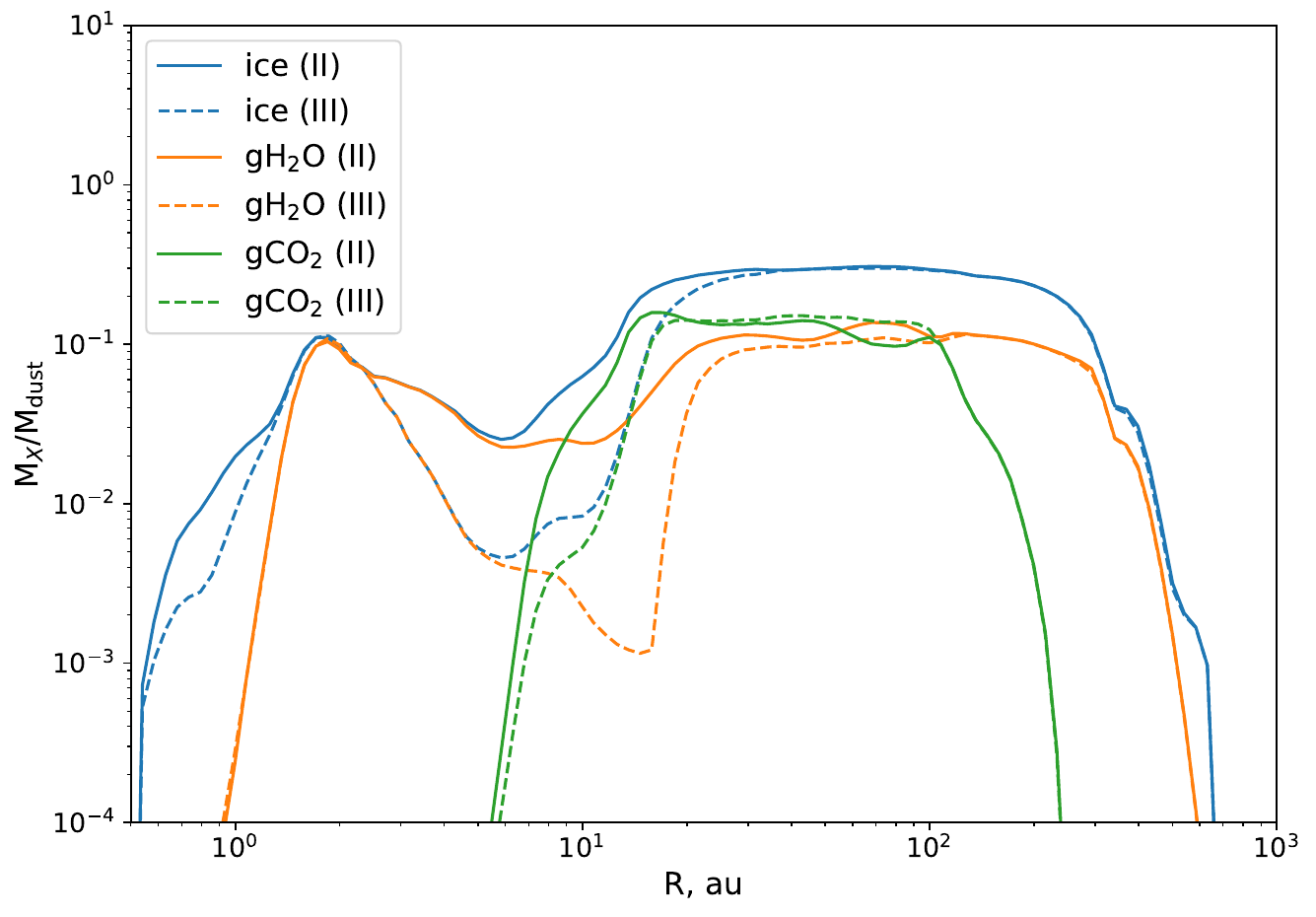}
}
\subfloat{
\includegraphics[width=0.5\textwidth]{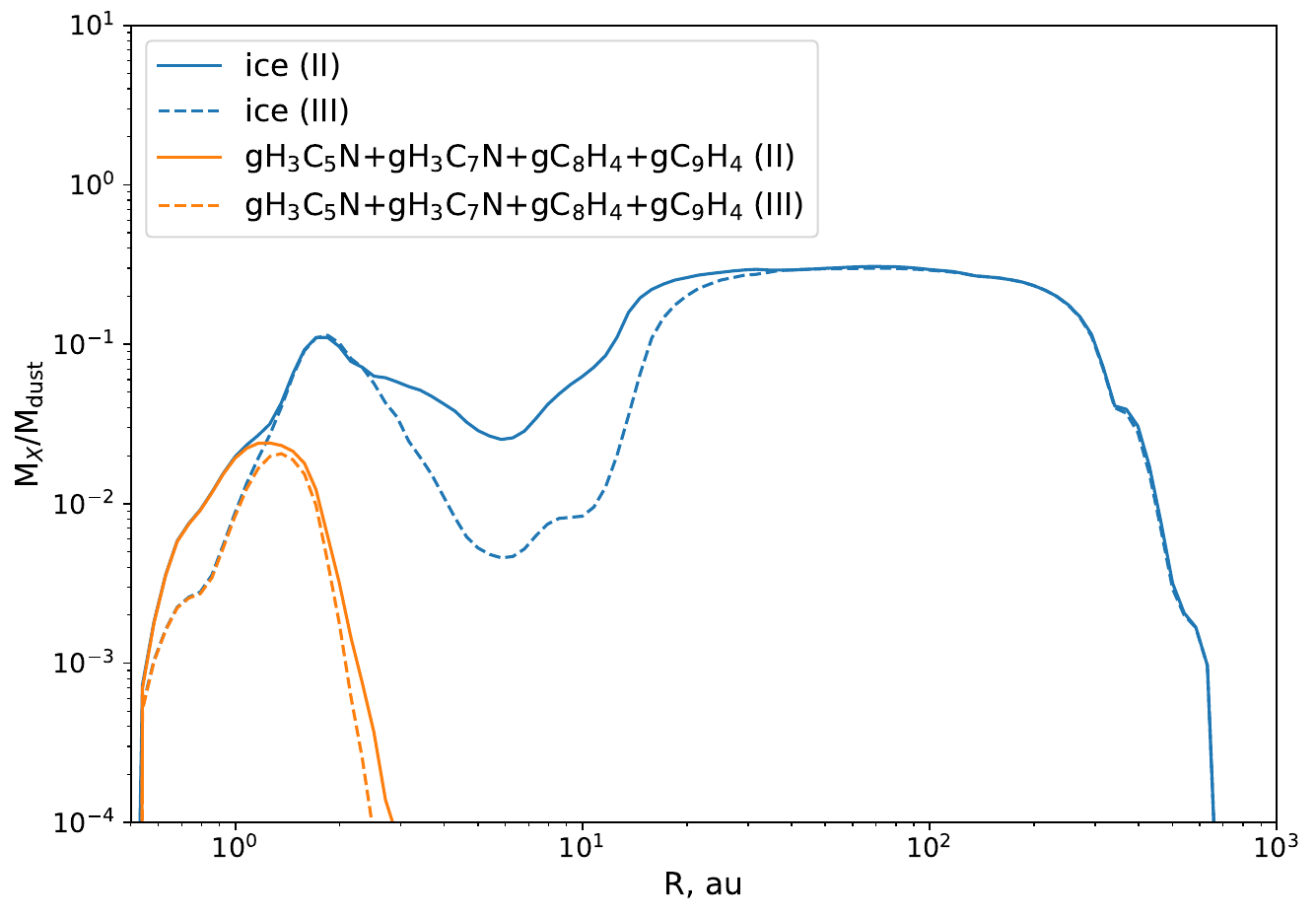}
}
\caption{Radial distribution of vertically integrated ice mass fraction in different models at the 1\,Myr time moment in models \Rmnum{2} and \Rmnum{3}. Left panel also shows gH$_2$O mass fraction and gCO$_2$ mass fraction. Right panel additionally shows a mass fraction of a sum of gH$_3$C$_5$N, gH$_3$C$_7$N, gC$_8$H$_4$ and gC$_9$H$_4$. Solid line is model \Rmnum{2} and dashed is model \Rmnum{3}.}
\label{fig:icemass2}
\end{figure}

Comparing now models \Rmnum{3} and \Rmnum{2}, we can see a significant drop in M$_{\rm ice}$/M$_{\rm dust}$ in the region between 2 and 20\,au (more than an order of magnitude at $R\approx6$\,au) and a smaller drop in the inner region closer than 1.35\,au. To identify their origin, we look at mass fractions of individual species. Fig.~\ref{fig:icemass2} shows such radial distributions in models \Rmnum{2} and \Rmnum{3}. According to left panel, we can identify that a drop in ice mass fraction is caused by a drop in gCO$_2$ mass fraction between 5 and 20\,au and gH$_2$O mass fraction between 2 and 20\,au. From the right panel we see that in the inner disc the decrease is caused by a change in the mass fractions of complex organic molecules gH$_3$C$_5$N, gH$_3$C$_7$N, gC$_8$H$_4$ and gC$_9$H$_4$.

\begin{figure}[h!]
\centering
\includegraphics[width=0.7\textwidth]{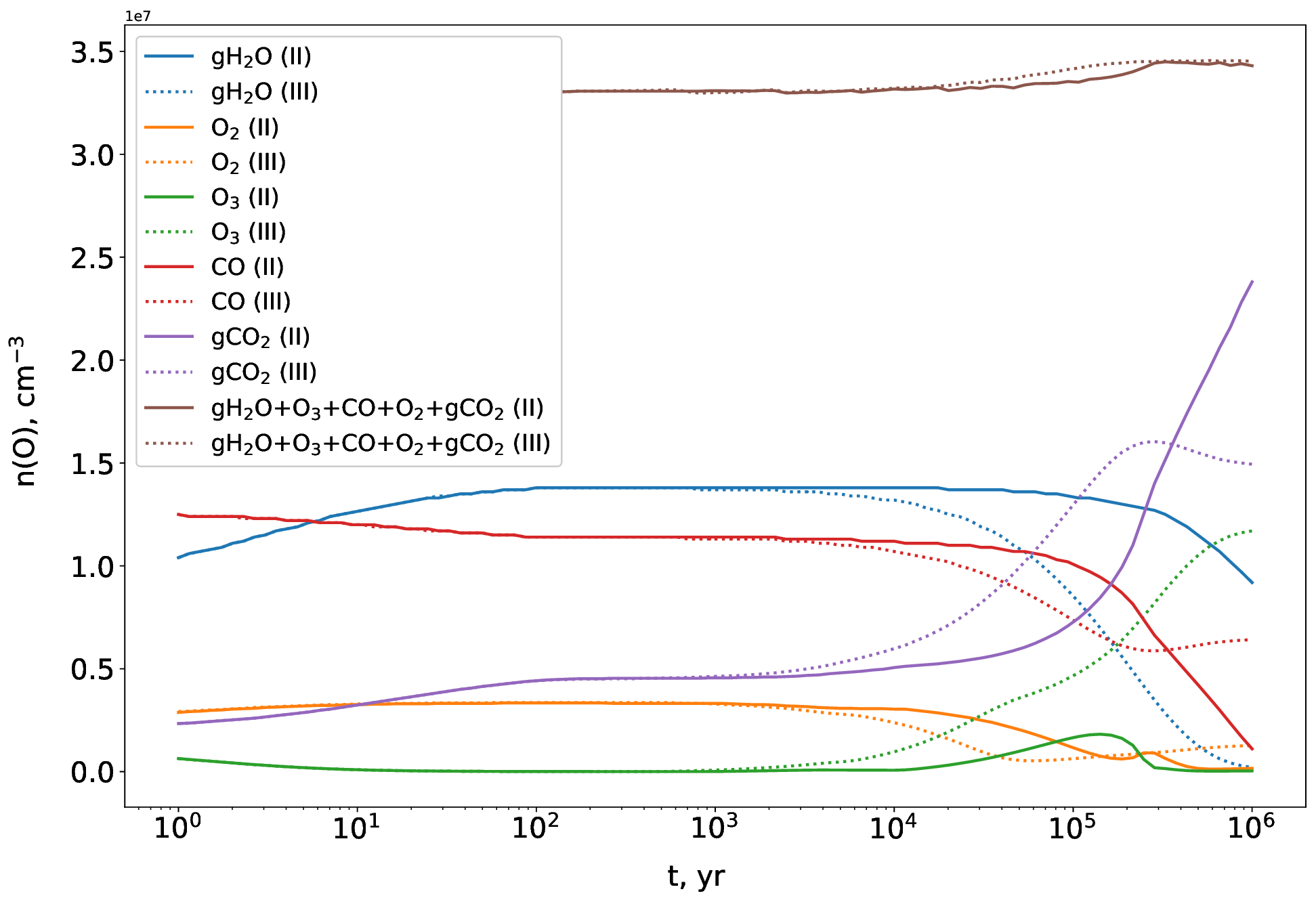}
\caption{Evolution of the number density of oxygen atoms contained in different species at $R \approx 15$\,au in the disc midplane. Colour denotes different species, solid lines represent model \Rmnum{2}, dotted lines represent model \Rmnum{3}.}
\label{fig:odens}
\end{figure}

To see what chain of reactions causes the decrease in water ice, we look at the time evolution of the number density of oxygen atoms contained in different species. Fig.~\ref{fig:odens} shows this evolution for the selected species and in total at 15\,au where the local minimum of gH$_2$O mass fraction is located. We can see that the total amount of oxygen contained in gH$_2$O, CO, O$_3$, O$_2$ and gCO$_2$ does not change with time and is roughly the same in models \Rmnum{2} and \Rmnum{3}. This indicates that the reactions responsible for the decrease in water ice happen between these species.

We identify the reactions responsible for the conversion between these species. Water ice gH$_2$O is photodissociated into gOH and gH by CRPs. Due to increased CRP-reaction rates in model \Rmnum{3}, this dissociation is quicker and becomes noticeable earlier than in model \Rmnum{2} (after 1000\,yr in model \Rmnum{3} vs $10^5$\,yr in model \Rmnum{2}). Released gOH reacts with CO, which at these temperatures only briefly stays on dust, to form gCO$_2$ and gH. It is reflected in Fig.~\ref{fig:odens} as anti-symmetric evolution of CO and gCO$_2$ number density.

At $10^5$\,yr, water ice in model \Rmnum{3} becomes too depleted to support the CO depletion reaction further and the trends in CO and gCO$_2$ number densities reverse.

Gas-phase ozone O$_3$ is also a part of the invariant total; it forms on dust in the gO+gO$_2$ reaction. Ice-phase atomic oxygen gO is formed in many CRP-reactions but the main among them is gCO$_2$ photodissociation. This connects the accelerated rise of O$_3$ number density with the gH$_2$O destruction in Fig.~\ref{fig:odens}. 

A slight rise in O$_2$ in model \Rmnum{3} at $10^5$\,yr comes from O$_3$ CRP-photodissociation that becomes more efficient because of the increased number density of O$_3$. In model \Rmnum{2}, the described process starts only around $10^5$\,yr. 

Concluding, an increase in CRP-reaction rates leads to a significantly faster photodissociation of water ice. Number densities of water ice in models \Rmnum{3} and \Rmnum{2} differ by an order of magnitude at 1\,Myr. Ice mass fraction at 15 au, however, is not significantly different, as it is dominated by gCO$_2$. Nonetheless, this change is still described by the chain of reactions above.

\section*{6. Conclusions}

In this work we explore the impact of the increase in the rates of CRP-reactions due to dust growth through two generalised chemical characteristics: ionisation fraction and ice mass fraction.

We show that the increase in the rates of CRP-reactions significantly changes ice mass fraction in the region between 2 and 20\,au due to an increased CRP-photodissociation of water ice. Ice mass fraction decreases there by an order of magnitude specifically because of a change in CRP-reaction rates. We note that in general, the characteristic effect of increasing the rate of CRP-reactions is the acceleration and earlier onset of processes involving these reactions. The dust growth itself causes a shift of ice mass fraction distribution away from the star as snowline positions depend on the dust size. There is also a decrease in ice mass fraction at the disc periphery due to lower dust opacity.

We also show that CRPs have very little impact on ionisation fraction in the disc. An effect from dust growth is more significant as it changes disc opacity to radiation and how efficiently grains are charged by electrons.

An increase in CRP-reaction rates is considered in~\cite{2012A&A...537A.138C} as well, where they note a similar drop in water ice number density with an increase in CO$_2$ ice around $\sim 10$\,au. However, this work does not consider surface chemistry beyond CO, CO$_2$, CH$_4$ and H$_2$O, so our results are mostly different from their findings.

Concluding, CRP-reactions have a noticeable effect on the chemical composition of the protoplanetary disc. We recommend the community to be cautious of the CRP-reaction rate values for more precise astrochemical modelling results. Recalculation of these rates considering dust evolution in protoplanetary discs is a relevant task to our field.

The work was supported by the Foundation for the Advancement of Theoretical Physics and Mathematics ``BASIS'' (20-1-2-20-1).

\bibliographystyle{inasan}
\small
\bibliography{References}
\normalsize

\end{document}